\documentclass[eqsecnum,showpacs,pre,twocolumn]{revtex4}

\usepackage{graphicx}
\usepackage{amsmath}
\usepackage{bm}
\usepackage{epsfig}
\usepackage{amsfonts}
\usepackage{amssymb}

\begin{document}

\hyphenation{Son-der-forsch-ungs-be-reich}

\title{Logarithmic Corrections in Dynamic Isotropic Percolation}

\author{Hans-Karl Janssen}
\affiliation{Institut f\"{u}r Theoretische Physik III, 
Heinrich-Heine-Universit\"{a}t
40225 D\"{u}sseldorf, Germany}

\author{Olaf Stenull}
\affiliation{Department of Physics and Astronomy, University of Pennsylvania,
Philadelphia PA 19104, USA}

\date{\today}

\begin{abstract}
\noindent Based on the field theoretic formulation of the general epidemic
process we study logarithmic corrections to scaling in dynamic isotropic
percolation at the upper critical dimension $d=6$. Employing renormalization
group methods we determine these corrections for some of the most interesting
time dependent observables in dynamic percolation at the critical point up to
and including the next to leading correction. For clusters emanating from a
local seed at the origin we calculate the number of active sites, the radius of gyration as well as
the survival probability.

\end{abstract}
\maketitle

\hyphenation{Son-der-for-schungs-be-reich}

\section{Introduction}
\noindent
Spreading phenomena occur in nature in many kinds with examples ranging from 
epidemics or forest fires \cite{BrHa57,Mo77,Li85,Du88,Ba85} over the growth of
populations \cite{Ba85,Mu89} and the activity of catalyzers 
\cite{ZiGuBa86,JeFoDi90} to the formation of stars and galaxies
\cite{SchuSe78}. In general terms the spreading of nonconserved agents has the 
following scenario: An agent (e.g., an active site of a lattice, an infected 
individual or a burning tree) randomly activates one or more of its neighbors. 
In the next timestep these infected neighbors act as agents themselves and so 
on. Via this elementary reaction, the activation spreads out diffusively in 
$d$-dimensional space. Competition between the agents for the resources of new 
activations limits their local density. Moreover, an agent become
spontaneously deactivated after some time. The long term behavior of the process, assumed to be 
emanating from a seed or germ at the origin, depends crucially on the difference 
$\tau$ between the deactivation and the activation rate. For $\tau>\tau_{c}$ the
process spreads, approaching a homogeneous steady state, over the entire space. 
For $\tau<\tau_{c}$ the spreading will finally ebb away as an inactive extinct 
state is approached (with possibly a static disturbance of the initial state, 
but only in a finite volume). The critical point $\tau=\tau_{c}$ separates this 
endemic absorbing phase from the epidemic active phase. Spreading near 
$\tau=\tau_{c}$ constitutes a critical phenomenon and is described by universal 
scaling laws.

There exist two fundamental universality classes of critical spreading
phenomena depending on the nature of the debris of the elementary
deactivation process. If the deactivated agents can recover so that they may 
become newly activated (so called simple epidemic),
the process belongs to the directed percolation (DP) universality class 
\cite{Ja81} (for a
review see, e.g., Ref.~\cite{Hi01}). Here, in the active phase, one can have
an epidemic surviving \textit{in loco}. On the other hand, if the debris
stays inactive forever (so called general epidemic), the spreading process 
becomes locally extinct. In this case, the process belongs to the dynamic 
isotropic percolation (dIP) universality class and the statistics of
the clusters formed by the debris are described by the usual percolation
theory (for reviews see, e.g., Refs.~\cite{StAh94,BuHa96}). Here, the
epidemic cannot, of course, survive \textit{in loco}, but an infinite epidemic
is nevertheless possible in the form of a solitary wave of activity. When
starting from a punctual seed, this leads to annular growth (e.g., fairy rings 
in two dimensions).

Most significantly, perhaps, numerical simulations and renormalization group
methods have contributed to our present understanding of spreading phenomena.
Until recently, accurate numerical investigations have been limited to lower
spatial dimensions. However, due to the staggering pace of hardware
improvements and the development of sophisticated algorithms, numerical
results for spreading in high dimensions become available
\cite{PaZiSt01,Gra02,osterkamp&Co_2003,Gra_pr,Lue03}. Some of these
results~\cite{Gra_pr,Lue03} clearly indicate the importance of
logarithmic corrections to scaling in the upper critical dimensions $d=4$ for
DP and $d=6$ for dIP. Analytic work on logarithmic corrections to percolation started not long after renormalization group methods became available. Essam {\em et al.}~\cite{essam&Co_78} determined the logarithmic corrections for the probability $P_\infty$ of belonging to an infinite cluster, the clusters' mean-square size $S$ and the correlation length $\xi$ as functions of the deviation from criticality $\tau - \tau_c$ as well as the logarithmic corrections at criticality to a "ghost" field $H$ as a function of $P_\infty$. Aharony~\cite{aharony_80} investigated logarithmic corrections in the context of universal amplitude ratios. More recently, Ruiz-Lorenzo~\cite{ruiz-lorenzo_98} presented the logarithmic corrections to  $S$ and $\xi$ at criticality as functions of the system size. In an upshot one can say that the previous results on logarithmic corrections in percolation are (i) restricted to static percolation, and (ii) limited to the leading correction.

For another system prominent in statistical physics,
viz.\ linear polymers, logarithmic corrections observed in simulations have
been very successfully described by renormalized field
theory~\cite{GHS94,GHS99}. It turned out, however, that the leading logartihmic corrections are not sufficient to yield a satisfactory agreement with the numerical data. To the contrary, it was found that it is crucial to include the next to leading correction. We expect a similar importance of the second order correction in the percolation problem.

The goal of this paper is to present analytical results for dIP in $d=6$ that are, with reasonable expectation, accurate enough to to yield a satisfactory agreement with numerical simulations. We derive these results for some of the most interesting observables in
dynamic percolation, namely the number $N(t)$ of agents (active infected sites) at
time $t$ generated by a seed at the space- and time-wise origin $(\mathbf{x}
=\mathbf{0},t=0)$, the survival probability $P(t)$ of the corresponding
cluster as well as the mean distance $R(t)$ of the agents from the
origin (radius of gyration).

The logarithmic corrections that we study here will not be found in real physical systems because the upper critical dimension 6 for dIP does not coincide with physical dimensions, $d=2$ or $d=3$. However, our results are sure to be valuable with respect to 
numerical simulations. Our final analytic expressions are well suitable for 
comparison to numerical data. Moreover, our results define a non-universal time 
scale that signals the onset of asymptotic behavior. This time scale may be used 
to assess the effective significance of a given microscopic simulation model for 
the dIP universality class. 

Complementary to the work presented here we have investigated logarithmic 
corrections in DP. Our results on observables akin to the quantities studied 
here are certainly equally interesting and will be presented in the near
future~\cite{JaSt_u1}. Taking a third route, we explored logarithmic corrections 
in static IP with emphasis on transport properties. A paper on this 
subject~\cite{StJa03} will be available soon.

The outline of the present paper is the following: In Sec.~\ref{reviewGEP} we
briefly review the renormalized field theory of the general epidemic process
(GEP) and previous renormalization group results on dIP. Moreover, we
conduct some general considerations about logarithmic corrections in the given
context. Section~\ref{logCorr} hosts the core of our analysis and contains the
main results. Section~\ref{conclus} concludes the main part of this paper with a discussion of our results and several remarks. Details of our diagrammatic perturbation calculation are relegated to an Appendix.

\section{Renormalized field theory of the GEP -- a brief review and general
considerations on logarithmic corrections}

\label{reviewGEP} In this section we briefly review the field theoretic 
description of the GEP and its renormalization. The aim is to provide the reader 
with background and to establish notation as
well as known results that we need as we go along. Furthermore, we outline the 
general structure of the sought after logarithmic corrections.

The GEP \cite{Mo77,Ba85,Mu89} is a kinetic growth model that was introduced by
Grassberger \cite{Gra83} as a lattice model of dIP. On a coarse grained scale,
the GEP can be minimally formulated by means of the Langevin equation in the Ito 
sense \cite{Ja85}
\begin{subequations}
\label{langAndNoise}%
\begin{align}
\lambda^{-1}\dot{n}(\mathbf{x},t)  &  =\nabla^{2}n(\mathbf{x},t)-\tau
n(\mathbf{x},t)-gn(\mathbf{x},t)\nonumber\\
&  \times \lambda \int_{-\infty}^{t}dt^{\prime}\,n(\mathbf{x},t^{\prime})+\zeta
(\mathbf{x},t)\,,\label{LangEq}\\
\overline{\zeta(\mathbf{x},t)\zeta(\mathbf{x}^{\prime},t^{\prime})}  &
=\lambda^{-1}g^{\prime}n(\mathbf{x},t)\delta(t-t^{\prime})\delta
(\mathbf{x}-\mathbf{x}^{\prime})\,. \label{Noise}%
\end{align}
Here $n(\mathbf{x},t)$ is the density of infected particles at time $t$ and
space coordinate $\mathbf{x}$. The variable $\tau$ is essentially the rate 
difference mentioned in the introduction (shifted by $\tau_c$) and specifies the deviation
from criticality. $\lambda$ represents a kinetic coefficient. The Gaussian
random field $\zeta(\mathbf{x},t)$ subsumes reaction noise and otherwise
neglected microscopic details (the overbar indicates averaging over its
distribution). Its correlation respects the existence of the absorbing state.
Many other analytic terms are conceivable as contributing to
Eqs.~(\ref{langAndNoise}). These, however, turn out to be irrelevant in the
sense of the renormalization group. Especially, a diffusional noise
contribution, relevant for diffusion limited reactions with multiplicative
noise, can be neglected.

Langevin equations are only a convenient shorthand for a stochastic process.
For the application of renormalized field theory, however, a path integral 
formulation
of the GEP is more adequate than the Langevin equation~(\ref{langAndNoise}).
The dynamic functional~\cite{Ja76,DeDo76,Ja92}, or in a more recent
terminology the response functional, of the GEP is given
by~\cite{Ja85,CaGra85}
\end{subequations}
\begin{equation}
\mathcal{J}=\int d^{d}x\,dt\,\lambda\tilde{s}\Big(\lambda^{-1}\frac{\partial
}{\partial t}+(\tau-\nabla^{2})+g(S-\frac{1}{2}\tilde{s})\Big)s\,.
\label{Funkt}%
\end{equation}
In deriving $\mathcal{J}$ from the Langevin equation one exploits a rescaling
form invariance of the response functional $\mathcal{J}$ that allows to equate
$g$ and $g^{\prime}$. $s(\mathbf{x},t)$ is proportional to $n(\mathbf{x},t)$.
$\tilde{s}(\mathbf{x},t)$ is the response field corresponding to
$s(\mathbf{x},t)$. The $S$ in the functional~(\ref{Funkt}) stands for the 
density of debris
$S(\mathbf{x},t)=\lambda\int_{-\infty}^{t}dt^{\prime}\,s(\mathbf{x},t^{\prime
})$. Note that
\begin{equation}
\tilde{s}(\mathbf{x},t)\longleftrightarrow-S(\mathbf{x},-t) \label{Zeitsp.}%
\end{equation}
is a symmetry transformation of the response functional~\cite{Ja85}.

$\mathcal{J}$ presents a vantage point for a systematic perturbation
calculation in the coupling constant $g$. Most economically this calculation
can be done by using dimensional regularization and minimal subtraction. Using
this scheme, the critical point value $\tau=\tau_{c}$ is formally set to zero
by the perturbational expansion. Generally, $\tau_{c}$ is a non-analytical
function of the coupling constant $g$. Thus, we implicitly make the additive
renormalization $\tau-\tau_{c}\rightarrow\tau$. For background on these
methods we refer to Refs. \cite{Am84,ZJ96}. An appropriate renormalization
scheme is
\begin{subequations}
\label{RenSch}%
\begin{align}
s\rightarrow\mathring{s}=Z^{1/2}s\,,  &  \quad\tilde{s}\rightarrow
\mathring{\tilde{s}}=\tilde{Z}^{1/2}\tilde{s}\,,\\
\lambda\rightarrow\mathring{\lambda}=Z^{-1/2}\tilde{Z}^{1/2}\lambda\,,  &
\quad\tau\rightarrow\mathring{\tau}=\tilde{Z}^{-1}Z_{\tau}\tau\,,\\
g\rightarrow\mathring{g}=\tilde{Z}^{-3/2}Z_{u}^{1/2}g\,,  &  \quad
G_{\varepsilon}g^{2}=u\mu^{\varepsilon}\,.
\end{align}
\end{subequations}
Here, the $\mathring{}$ indicates unrenormalized quantities. $\varepsilon=6-d$
measures the deviation from the upper critical dimension. The factor
$G_{\varepsilon}=\Gamma(1+\varepsilon/2)/(4\pi)^{d/2}$ is introduced
exclusively for later convenience. $\mu$ is an external inverse length scale.
Note that the renormalizations~(\ref{RenSch}) preserve the
invariance~(\ref{Zeitsp.}). The renormalization factors $\tilde{Z}$, $Z_{\tau}$, and
$Z_{u}$ are known to 3-loop order~\cite{AKM81}. One of us~\cite{Ja85}
calculated $Z$ to 2-loop order. In the following we will need the
renormalization factors explicitly to 1-loop order to which they are given by
\begin{subequations}
\label{Z-Fakt}%
\begin{align}
\tilde{Z}=1+\frac{u}{6\varepsilon}+\cdots\,,  &  \quad Z=1+\frac
{4u}{3\varepsilon}+\cdots\,,\\
Z_{\tau}=1+\frac{u}{\varepsilon}+\cdots\,,  &  \quad Z_{u}=1+\frac
{4u}{\varepsilon}+\cdots\,.
\end{align}

The critical behavior of the Green's functions $G_{n,\tilde{n}}=\langle\lbrack
s]^{n}[\tilde{s}]^{\tilde{n}}\rangle^{(cum)}$ is governed by the
Gell-Mann--Low renormalization group equation (RGE)
\end{subequations}
\begin{equation}
\Big[\mathcal{D}_{\mu}+\frac{1}{2}\bigl(n\gamma+\tilde{n}\widetilde{\gamma
}\bigr)\Big]G_{n,\tilde{n}}(\{\mathbf{r},t\};\tau,u;\lambda,\mu)=0 \label{RGG}%
\end{equation}
with the differential operator
\begin{equation}
\mathcal{D}_{\mu}=\mu\partial_{\mu}+\lambda\zeta\partial_{\lambda}+\tau
\kappa\partial_{\tau}+\beta\partial_{u}\,. \label{RGOp}%
\end{equation}
The Wilson functions appearing in the RGE are given to 2-loop order by
\cite{AKM81,Ja85}
\begin{subequations}
\label{RGFunkt}%
\begin{align}
\gamma &  =-\frac{4}{3}u+\Big(\frac{1895}{54}+9\ln3-5\ln4\Big)\frac{u^{2}}%
{16}\,,\\
\tilde{\gamma}  &  =-\frac{1}{6}u+\frac{37}{216}u^{2}\,,\\
\kappa &  =\frac{5}{6}u-\frac{193}{108}u^{2}\,,\\
\zeta &  =\frac{\gamma-\tilde{\gamma}}{2}=-\frac{7}{12}u+\Big(\frac{1747}%
{54}+9\ln3-5\ln4\Big)\frac{u^{2}}{32}\,,\\
\beta &  =-\varepsilon u+\frac{7}{2}u^{2}-\frac{671}{72}u^{3}+\Big(\frac
{414031}{2592}+93\zeta(3)\Big)\frac{3u^{4}}{16}\,.
\end{align}
Note that we have stated $\beta$ to 3-loop order because the high order contribution improves our quantitative predictions. In the
remainder we will adopt a convenient abbreviated notation for the Wilson
functions of the type $f(u)=f_{0}+f_{1}u+f_{2}u^{2}+\cdots$ with $f$ standing
for $\gamma$, $\tilde{\gamma}$, $\kappa$, $\zeta$, and $\beta$, respectively.
The meaning of the coefficients $f_{0}$, $f_{1}$ and so on should be evident.

The RGE can be solved by the method of characteristics. To this end one
introduces a flow parameter $l$ and sets up characteristic equations that
describe how the scaling parameters transform if the external momentum scale
is changed by varying $\bar{\mu}(l)=\mu l$. The characteristic for the
dimensionless coupling constant $u$ reads
\end{subequations}
\begin{equation}
l\frac{dw}{dl}=\beta(w)\, \label{Char-u}%
\end{equation}
where we abbreviated $w=\bar{u}(l)$. Solving this differential equation for
$\varepsilon=6-d=0$ yields
\begin{equation}
l=l(w)=l_{0}w^{-\beta_{3}/\beta_{2}^{2}}\exp\bigg[-\frac{1}{\beta_{2}w}%
+\frac{(\beta_{3}^{2}-\beta_{2}\beta_{4})}{\beta_{2}^{3}}w+O(w^{2})\bigg]\,,
\label{l(w)}%
\end{equation}
where $l_{0}$ is an integration constant. The remaining characteristics are
all of the same structure, namely
\begin{equation}
l\frac{d\ln\bar{Q}(w)}{dl}=q(w)\,.
\end{equation}
Here, $Q$ is a placeholder for $Z$, $\tilde{Z}$, $Z_{\tau}$, and $Z_{\lambda}%
$, respectively. $q$ is an ambiguous letter for $\gamma$, $\tilde{\gamma}$,
$\kappa$, and $\zeta$, respectively. Exploiting $ld/dl=\beta d/dw$ we obtain
the solution
\begin{equation}
\bar{Q}(w)=Q_{0}w^{q_{1}/\beta_{2}}\exp\bigg[\frac{(q_{2}\beta_{2}-q_{1}%
\beta_{3})}{\beta_{2}^{2}}w+O(w^{2})\bigg]\,, \label{Q(w)}%
\end{equation}
where $Q_{0}$ symbolizes a non universal integration constant.

With the solutions to the characteristics the scaling behavior of the Green's
functions is found to be
\begin{align}
&  G_{n,\tilde{n}}(\{\mathbf{x},t\};\tau,u;\lambda,\mu
)\nonumber\label{GrFuSkal}\\
&  =\,(\mu l)^{n(d+2)/2+\tilde{n}(d-2)/2}Z(w)^{n/2}\tilde{Z}(w)^{\tilde{n}%
/2}\\
&  \times\,G_{n,\tilde{n}}(\{l\mu\mathbf{x},Z_{\lambda}(w)(l\mu)^{2}\lambda
t\};Z_{\tau}(w)\tau/(\mu l)^{2},w;1,1)\,.\nonumber
\end{align}
The flow parameter introduced via the characteristics is arbitrary. Thus, we
have a freedom of choice that can be exploited to re-scale the relevant
variables, viz. $\mathbf{x}$, $t$ and $\tau^{-1}$, so that they acquire a
finite asymptotic value. For the goals pursuit in this paper, an appropriate
choice is
\begin{equation}
Z_{\lambda}(w)(l\mu)^{2}\lambda t=X_{0}\,, \label{Wahl_X}%
\end{equation}
where $X_{0}$ is a constant of order unity. With this choice $w$ and $l$ tend
to zero for $\lambda\mu^{2}t\rightarrow\infty$. Based on our 
choice~(\ref{Wahl_X}) we introduce the convenient time variable
\begin{equation}
s=\frac{\beta_{2}}{2}\ln\bigl(t/t_{0}\bigr)=\frac{7}{4}\ln\bigl(t/t_{0}%
\bigr) \, , 
\label{s(t)}
\end{equation}
where $t_{0}\propto X_{0}$ is a non universal time constant. From Eq.~(\ref{l(w)})
and Eq.~(\ref{Q(w)}), specialized to $Z_{\lambda}$, we get
\begin{equation}
s=w^{-1}-a_{1}\ln w+a_{2}w+O(w^{2}) \label{t(w)}%
\end{equation}
for the derived time variable. The
constants $a_{1}$ and $a_{2}$ are given by
\begin{subequations}
\begin{align}
a_{1}  &  =\frac{\beta_{2}\zeta_{1}-2\beta_{3}}{2\beta_{2}}=\frac{1195}%
{504}=2.37103\,,\\
a_{2}  &  =\frac{\zeta_{1}\beta_{3}-\zeta_{2}\beta_{2}}{2\beta_{2}}%
+\frac{\beta_{2}\beta_{4}-\beta_{3}^{2}}{\beta_{2}^{2}}\\
&  =\frac{1766273}{1016064}+\frac{10\ln2-9\ln3}{64}+\frac{279\zeta(3)}%
{56}=7.68098\,.
\end{align}
Using Eq.\ (\ref{t(w)}) we obtain for the dimensionless coupling constant as a
function of time the expression
\end{subequations}
\begin{equation}
w=s^{-1}\exp\bigg[a_{1}\frac{\ln s}{s}+O\Big(\frac{\ln^{2}s}{s^{2}},\frac{\ln
s}{s^{2}},\frac{1}{s^{2}}\Big)\bigg]\,. \label{w(s)}%
\end{equation}

Exploiting Eqs.~(\ref{GrFuSkal}), (\ref{l(w)}), (\ref{Q(w)}) and (\ref{w(s)})
we find that the observables to be considered are of the form
\begin{equation}
\mathcal{A}=\mathcal{A}_{0}\exp\bigg[as+b\ln s+\frac{c\ln s+c^{\prime}}%
{s}+O\Big(\frac{\ln^{2}s}{s^{2}},\frac{\ln s}{s^{2}},\frac{1}{s^{2}%
}\Big)\bigg]\,. \label{Allg.G}%
\end{equation}
$\mathcal{A}_{0}$ is, like $t_{0}$, a non universal constant. $a$, $b$, $c$
and $c^{\prime}$ are universal numbers. $a$ stems from mean field theory. $b $
and $c$ represent 1- and 2-loop renormalization group results, respectively.
$c^{\prime}$ comprises contributions from the Wilson functions to 2-loop order
as well as an amplitude to be determined in an explicit 1-loop calculation of
$\mathcal{A}$. This amplitude depends on $X_{0}$, as do $s$ and 
$\mathcal{A}_{0}$. Over all, a variation of $X_{0}$ leaves $c^{\prime}$ invariant.

\section{Logarithmic corrections for the observables of interest}

\label{logCorr} Equipped with important intermediate results as well as some
knowledge of the structure of the Green's functions, we next determine the
sought after logarithmic corrections. Since we already know the general form of the results, this part will be fairly brief. 

\subsection{Number of active particles}

The number of active particles generated by a seed at the origin is given at
criticality, $\tau=0$, by
\begin{align}
N(t)  &  =\int d^{d}x\,G_{1,1}(\mathbf{x},t;0,u;\lambda,\mu)\nonumber\\
&  =\bigl(Z(w)\tilde{Z}(w)\bigr)^{1/2}\int d^{d}x\,(\mu l)^{d}\nonumber\\
&  \times G_{1,1}(l\mu\mathbf{r},Z_{\lambda}(w)(l\mu)^{2}\lambda
t;0,w;1,1)\label{N(t)1}\\
&  =\bigl(Z(w)\tilde{Z}(w)\bigr)^{1/2}G_{1,1}(\mathbf{q}=0,X_{0}%
;0,w;1,1)\,.\nonumber
\end{align}
Specializing solution~(\ref{Q(w)}) to $Q=Z$ and $Q=\tilde{Z}$ we obtain
\begin{align}
&  \bigl(Z(w)\tilde{Z}(w)\bigr)^{1/2}\propto\exp\bigg[\frac{(\gamma_{1}%
+\tilde{\gamma}_{1})}{2\beta_{2}}\ln w\nonumber\\
&  +\,\bigg(\frac{(\gamma_{2}+\tilde{\gamma}_{2})\beta_{2}-(\gamma_{1}%
+\tilde{\gamma}_{1})\beta_{3}}{2\beta_{2}^{2}}\bigg)w+O(w^{2})\bigg]\,.
\label{ZZs}%
\end{align}
A perturbation expansion of the Green's function $G_{1,1}$ brings about an
amplitude $A_{N}(X_{0})$ that we define via
\begin{equation}
G_{1,1}(\mathbf{q}=0,X_{0};0,w;1,1)\propto\bigl(1+A_{N}(X_{0})w+O(w^{2}%
)\bigr)\,. \label{A_N}%
\end{equation}
This amplitude follows from our 1-loop calculation presented in
Appendix~\ref{app:G11} as
\begin{equation}
A_{N}(X_{0})=\frac{3}{8}\Big(\mathcal{Z}+\frac{3}{2}-\frac{\ln2}%
{3}\Big) \label{A_N(X)}%
\end{equation}
where we used the shorthand notation $\mathcal{Z}=\ln X_{0}+C_{E}$ with
$C_{E}$ being Euler's constant. Collecting Eqs.~(\ref{N(t)1}), (\ref{ZZs}) and
(\ref{A_N}) we find
\begin{align}
N(t)  &  =N_{0}\bigl(w^{-1}+B_{N}\bigr)^{a_{N}}\exp\bigl(c_{N}w+O(w^{2}
)\bigr)\nonumber\\
&  =N_{0}^{\prime}\bigl(s+B_{N}\bigr)^{a_{N}}\Big[1+\frac{b_{N}\ln s+c_{N}}
{s}
\nonumber \\
& +O\Big(\frac{\ln^{2}s}{s^{2}},\frac{\ln s}{s^{2}},\frac{1}{s^{2}
}\Big)\Big] \label{Npara1}
\end{align}
where $N_{0}$ is a non universal constant, $N_{0}^{\prime}$ is a non universal
constant slightly different from $N_{0}$, and $B_N = A_N/a_N$. The first row of
(\ref{Npara1}) and
the result (\ref{t(w)}) constitute a parametric representation of the tuple
$(N,s)$ that is suitable for comparison to numerical simulations. The second row of
(\ref{Npara1}) shows the more traditional form. The constants $a_{N}$, $b_{N}%
$, $c_{N}$, and $B_{N}$ are given by
\begin{subequations}
\begin{align}
\label{resan}
a_{N}  &  =-\frac{\gamma_{1}+\tilde{\gamma}_{1}}{2\beta_{2}}=\frac{3}%
{14}\,=0.214286\,,\\
b_{N}  &  =a_{N}\frac{2\beta_{3}-\zeta_{1}\beta_{2}}{2\beta_{2}}=-\frac
{1195}{2352}=-0.508078\,,\\
c_{N}  &  =\frac{\gamma_{2}+\tilde{\gamma}_{2}}{2\beta_{2}}-\beta_{3}%
\frac{\gamma_{1}+\tilde{\gamma}_{1}}{2\beta_{2}^{2}}\nonumber\\
&  =-\frac{365}{1568}+\frac{9\ln3-5\ln4}{112}=-0.206387\,,\\
B_{N}  &  =\frac{7}{4}\bigl(\mathcal{Z}+\frac{3}{2}-\frac{\ln2}{3}%
\bigr)=1.75\mathcal{Z}+2.22066\,.
\end{align}
\end{subequations}
Note from Eqs.~(\ref{s(t)}) and (\ref{Npara1}) that the arbitrary 
constant $\mathcal{Z}$ could be eliminated by a rescaling of the nonuniversal time
constant $t_{0}$. This finding will also apply to the remaining results stated below.

At this point we would like to warn against attempts to deduce the logarithmic corrections for dynamic quantities from the logarithmic corrections calculated for static percolation. Grassberger~\cite{Gra02}, for example, exploited the results of Essam {\em et al}~\cite{essam&Co_78} via replacing $\tau$ by $1/t$ on the grounds that the critical exponent $\nu_t$ for the correlation time is one in mean-field theory. This reasoning leads to $N(t) \sim [\ln (t)]^{2/7}$~\cite{footnote}. From Eqs.~(\ref{Npara1}) and (\ref{resan}), however, we see that the correct result, to leading order, is $N(t) \sim [\ln (t/t_0)]^{3/14}$. By merely using the mean field relation between $\tau$ and $t$ one misses contributions to the leading logarithmic term stemming from renormalization factors including $Z$ [cf.~Eq.~(\ref{ZZs})]. Since $Z$ is absent in static percolation~\cite{footnote2}, one cannot deduce the logarithmic behavior of the dynamic quantity $N(t)$ from the known results for static percolation. Likewise, it should not be attempted to combine our dynamic results, e.g. those for $N(t)$ and $R(t)$, to obtain predictions on logarithmic corrections in static percolation.

\subsection{Radius of gyration}

The mean square distance from the origin of the active particles is defined
as
\begin{equation}
R(t)^{2}=\frac{\int d^{d}x\,\mathbf{x}^{2}\,G_{1,1}(\mathbf{x},t)}{2d\int
d^{d}x\,G_{1,1}(\mathbf{x},t)}=-\left.  \frac{\partial\ln G_{1,1}%
(\mathbf{q},t)}{\partial q^{2}}\right\vert _{\mathbf{q}=0}. \label{R-def}%
\end{equation}
From the scaling form~(\ref{GrFuSkal}) it follows for $\tau=0$ that
\begin{align}
&  \left.  \frac{\partial\ln G_{1,1}(\mathbf{q},t)}{\partial q^{2}}\right\vert
_{\mathbf{q}=0}\nonumber\\
&  =\,\left.  \frac{\partial\ln G_{1,1}((l\mu)^{-1}\mathbf{q},Z_{\lambda
}(w)(l\mu)^{2}\lambda t;0,w;1,1)}{\partial q^{2}}\right\vert _{\mathbf{q}%
=0}\nonumber\\
&  =\,(l\mu)^{-2}\left.  \frac{\partial\ln G_{1,1}(\mathbf{q},X_{0}%
;0,w;1,1)}{\partial q^{2}}\right\vert _{\mathbf{q}=0}\,.
\end{align}
Incorporating the solutions to the appropriate characteristics and the results
of Appendix~\ref{app:G11}
\begin{align}
&  -\left.  \frac{\partial}{\partial q^{2}}\ln G_{1,1}(\mathbf{q},\lambda
\mu^{2}t=X_{0};0,w;1,1)\right\vert _{\mathbf{q}=0}\nonumber\\
&  =X_{0}\bigl(1+A_{R}(X_{0})w+O(w^{2})\bigr)\,,
\end{align}
with%
\begin{equation}
A_{R}(X_{0})=\frac{7}{24}\Big(\mathcal{Z}-\frac{2}{3}-\frac{\ln2}{7}\Big)\,,
\label{A_R(X)}%
\end{equation}
we find%
\begin{align}
t^{-1}R^{2}  &  =R_{0}^{2}\bigl(w^{-1}+B_{R}\bigr)^{a_{R}}\exp\bigl(c_{R}%
w+O(w^{2})\bigr)\nonumber\\
&  =R_{0}^{\prime2}\bigl(s+B_{R}\bigr)^{a_{R}}\Big[1+\frac{b_{R}\ln s+c_{R}%
}{s}
\nonumber\\
&+O\Big(\frac{\ln^{2}s}{s^{2}},\frac{\ln s}{s^{2}},\frac{1}{s^{2}%
}\Big)\Big]\,.
\end{align}
with $R_{0}^{2}$ and $R_{0}^{\prime2}$ being non-universal amplitudes. Here
the constants $a_{R}$, $b_{R}$, $c_{R}$, and $B_{R} = A_R/a_R$ are given by
\begin{subequations}
\begin{align}
a_{R}  &  =-\frac{\zeta_{1}}{\beta_{2}}=\frac{1}{6}=0.166666\,,\\
b_{R}  &  =a_{R}\frac{2\beta_{3}-\zeta_{1}\beta_{2}}{2\beta_{2}}=-\frac
{1195}{3024}=-0.395172\,,\\
c_{R}  &  =\frac{\zeta_{2}}{\beta_{2}}-\frac{\zeta_{1}\beta_{3}}{\beta_{2}%
^{2}}\nonumber\\
&  =-\frac{937}{6048}+\frac{9\ln3-5\ln4}{112}=-0.128534\,,\\
B_{R}  &  =\frac{7}{4}\bigl(\mathcal{Z}-\frac{2}{3}-\frac{\ln2}{7}%
\bigr)=1.75\mathcal{Z}-1.33995\,.
\end{align}
\end{subequations}

\subsection{Survival probability}

As shown in Ref.~\cite{Ja02}, the survival probability of an active cluster
emanating from a seed at the origin is given by
\begin{align}
P(t)  &  =-\lim_{k\rightarrow\infty}\langle\mathrm{e}^{-k\mathcal{N}}\tilde
{s}(-t)\rangle
\label{Surv.P.}%
\end{align}
where $\mathcal{N}=\int d^{d}x\,s(\mathbf{x},0)$. For the purpose of actual calculations, it is convenient to rewrite Eq.~(\ref{Surv.P.}) as
\begin{eqnarray}
\label{newSurvProb}
P(t) &=& -\lim_{k\rightarrow\infty}\langle\tilde
{s}(-t)\rangle_k
\nonumber \\
&=&-G_{0,1}(\mathbf{-}t,\tau,k=\infty,u;\lambda,\mu)
\end{eqnarray}
where $\langle \cdots \rangle_k$ stands for averaging with respect to the response functional $\mathcal{J}_k$ that is obtained upon augmenting the original reponse functional~(\ref{Funkt}) by a source $k(t)=k\delta(t)$ conjugate to the field $s$:
\begin{eqnarray}
\mathcal{J}_k = \mathcal{J} + \int dt\, k(t)\mathcal{N}(t) \, .
\end{eqnarray}
With this source present, one no longer has
$\langle\tilde{s}\rangle=0$. To avoid tadpoles in our perturbation
calculation, we perform a shift $\tilde{s}\rightarrow\tilde{s}+\tilde{M}$ so
that $\langle\tilde{s}\rangle=0$ is restored This procedure leads to the new response functional
\begin{align}
&  \mathcal{J}_k=\int d^{d}x\,dt\,\biggl(\lambda
\tilde{s}\Big(\lambda^{-1}\frac{\partial}{\partial t}\nonumber\\
&  +\,(\tau-g\tilde{M}-\nabla^{2})+g(S-\frac{1}{2}\tilde{s})\Big)s\nonumber\\
&  +\lambda g\tilde{M}sS+\Big(-\overset{.}{\tilde{M}}+\lambda\tau\tilde
{M}-\frac{\lambda g}{2}\tilde{M}^{2}+k\Big)s\biggr)\,. \label{J_k}%
\end{align}
Based on this functional we calculate $G_{0,1} = \tilde{M}$ to 1-loop order. Some details of this calculation are in Appendix~\ref{app:G01}. We obtain
\begin{align}
&  G_{0,1}(\mathbf{-}X_{0},0,k=\infty,w;1,1)\nonumber\\
&  \propto\,w^{-1/2}\bigl(1+A_{P}(X_{0})w+O(w^{2})\bigr)\,. \label{A_P}%
\end{align}
with the amplitude $A_{P}(X_{0})$ reading
\begin{equation}
A_{P}(X_{0})=\frac{5}{8}\Big(\mathcal{Z}+1-\frac{11\ln2}{5}\Big)\,.
\label{A_P(X)}%
\end{equation}
Recalling the scaling form~(\ref{GrFuSkal}) and our choice for the flow parameter we deduce that, for $\tau=0$,
\begin{equation}
P(t)=-\tilde{Z}(w)^{1/2}(\mu l)^{2}G_{0,1}(\mathbf{-}X_{0},0,\infty,w;1,1)\,.
\end{equation}
Collecting, we then obtain
\begin{align}
tP(t)  &  =P_{0}\bigl(w^{-1}+B_{P}\bigr)^{a_{P}}\exp\bigl(c_{P}w+O(w^{2}%
)\bigr)\nonumber\label{Ppara}\\
&  =P_{0}^{\prime}\bigl(s+B_{P}\bigr)^{a_{P}}\Big[1+\frac{b_{P}\ln s+c_{P}}%
{s}
\nonumber \\
&+O\Big(\frac{\ln^{2}s}{s^{2}},\frac{\ln s}{s^{2}},\frac{1}{s^{2}%
}\Big)\Big]\,.
\end{align}
$P_{0}$ and $P_{0}^{\prime}$ are simply related non universal amplitudes. The
constants $a_{P}$, $b_{P}$, $c_{P}$, and $B_{P} = A_P/a_P$ are given by
\begin{subequations}
\begin{align}
a_{P}  &  =\frac{2\zeta_{1}+\beta_{2}-\tilde{\gamma}_{1}}{2\beta_{2}}=\frac
{5}{14}=0.357143\,,\\
b_{P}  &  =a_{P}\frac{2\beta_{3}-\zeta_{1}\beta_{2}}{2\beta_{2}}=-\frac
{5975}{7056}=-0.846797\,,\\
c_{P}  &  =\frac{\tilde{\gamma}_{2}-2\zeta_{2}}{2\beta_{2}}+\beta_{3}%
\frac{2\zeta_{1}-\tilde{\gamma}_{1}}{2\beta_{2}^{2}}\nonumber\\
&  =\frac{1637}{14112}-\frac{9\ln3-5\ln4}{112}=0.0896074\,,\\
B_{P}  &  =\frac{7}{4}\bigl(\mathcal{Z}+1-\frac{11\ln2}{5}%
\bigr)=1.75\mathcal{Z}-0.918617\,.
\end{align}
\end{subequations}

\section{Discussion of results and concluding remarks}
\label{conclus} 

As far as time dependent observables in percolation are concerned, we are not aware of any previous analytic work addressing logarithmic corrections. More general, we do not know of any work that has determined logarithmic corrections in percolation (static or dynamic, IP or DP) beyond the leading corrections. Here we went beyond the leading terms, and hence we are confident that our results compare well with simulations, perhaps even quantitatively. For linear polymers it turned out that the knowledge of the leading logarithmic correction is not sufficient for a good agreement between simulation data and theory. Rather, the next to leading corrections turned out to be crucial in comparing numerical and analytical results. We expect the same for percolation. Indeed, preliminary Monte Carlo results corroborate this expectation~\cite{Gra_pr,Lue03}.

Our results define a non-universal time-scale $t_{0}$. For times $t$ greater then $t_{0}$ we expect the validity of our asymptotic expansions. The time-scale $t_{0}$ can be utilized as a measure of quality for different microscopic models of dynamical percolation. Thus, our results may guide those performing simulations in choosing the most efficient model.

It is interesting to note that the time-scale $t_{0}$ has an analog in quantum chromodynamics. For times greater than $t_{0}$ the model becomes asymptotically free. Thus, with the exchange of an infrared(IR)-free theory to an ultraviolet(UV)-free theory,
$t_{0}$ corresponds to the hadronization scale of quantum chromodynamics. The
dependence of our results on this dimensional non-universal parameter $t_{0}$
parallels therefore the phenomenon of dimensional transmutation in
renormalizable asymptotically free quantum field theories that are naively
scale-free. 

Our results feature a mutual non-universal constant, viz. $\mathcal{Z}$. This constant could be eliminated by rescaling $t_0$. One might be tempted to think that one could eliminate the entire amplitudes, $A_N (X_0)$ and so on, from our results by rescaling $t_0$, and that the calculation of the amplitudes is hence superfluous. One has to keep in mind however, that one has to choose $t_0$ consistently for all observables. Thus, one cannot remove the amplitudes simultaneously from all the observables, and their calculation is indeed necessary. 

We refrain from eliminating $\mathcal{Z}$ from our results because it might be exploited, due to its non-universality, as a fit parameter. By fitting $\mathcal{Z}$ one can compensate partially for the effect of higher order terms that have been neglected in our calculations. In this sense one can think of $\mathcal{Z}$ as mimicking these higher order terms.  

When written as an explicit function of time the observables of interest have fairly complicated formulae. Using the parametric representation in terms of the effective coupling constant $w$ eases this situation. Moreover, the time and the observables posses a nicely systematic expansion in $w$ so that it is straightforward to keep track of the different orders in perturbation theory. In the traditional from, the orders are not so clear cut because nested functions of logarithms have to be compared. The parametric representations can be conveniently compared to simulations. Essentially, one just needs to make parametric plots of the tuples $(N,s)$, $(R,s)$ and $(P,s)$ and then compare the numerical data to these plots.

In order to improve the accuracy of our results, one needs a refined quantitative knowledge of the Wilson functions. Whereas the $\varepsilon$ expansion results for the critical percolation exponents have been improved by resummation techniques such as Pad\'e-Borel resummation~\cite{AKM81}, this kind of refinement has not yet been achieved for the percolation Wilson functions. Here lies an opportunity for useful future work for the ambitious. Another possibility for future work is to improve the results on static percolation mentioned in the introduction by calculating the next to leading logarithmic corrections. With the kind of field theoretic methods that we applied here, this is a reasonable task. 

Apparently, firm numerical results that are suitable for comparison to our analytical results are not available yet. We hope, however, that our work triggers increasing efforts in this direction, and that corresponding numerical results will be available in the near future.

\begin{acknowledgments}
This work has been supported by the Deutsche Forschungsgemeinschaft via the
Sonderforschungsbereich~237 ``Unordnung und gro{\ss }e Fluktuationen'' and the
Emmy Noether-Programm. We thank D. Stauffer for bringing our attention to several references.
\end{acknowledgments}

\appendix*

\section{Explicit calculation of Green's functions}

In this Appendix we outline our 1-loop calculation of scaling functions
belonging to the Green's functions $G_{1,1}$ and $G_{0,1}$. In particular, we
compute the amplitudes $A_{N}(X_{0})$, $A_{P}(X_{0})$, and $A_{R}(X_{0})$
entering the logarithmic corrections.

\subsection{The Green's function $G_{1,1}$}

\label{app:G11} 
\begin{figure}[ptb]
\includegraphics[width=2.75cm]{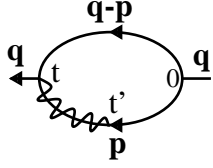}\caption{Self-energy $\Sigma
(\mathbf{q},t)$ at 1-loop order.}%
\label{selfEn}%
\end{figure}

A first step of any diagrammatic perturbation calculation is,
of course, the determination of the constituting elements. From the response
functional~(\ref{Funkt}) we gather the Gaussian propagator
\begin{equation}
G(\mathbf{q},t)=\theta(t)\, \exp\big[ -\lambda(\tau+q^{2})t\big] \label{Prop}%
\end{equation}
and the 3-leg vertices $\lambda g$ and $- \lambda^{2} g \, \theta(t
-t^{\prime}) $, where $\theta(t)$ denotes the step function. With these
elements, the self-energy $\Sigma(\mathbf{q},t)$ is given at 1-loop order by
the diagram depicted in Fig.~\ref{selfEn}.
This diagram stands for the mathematical formula,
\begin{equation}
\Sigma(\mathbf{q},t)=-\lambda^{3}g^{2}\int_{0}^{t}dt^{\prime}\int_{\mathbf{p}
}G(\mathbf{p},t^{\prime})G(\mathbf{q-p},t) \, . \label{Selbst}%
\end{equation}
After integrating out the loop momentum we can rewrite $\Sigma(\mathbf{q},t)$
as
\begin{align}
&  \Sigma(\mathbf{q},t)=-\frac{(\lambda g)^{2}}{(4\pi)^{d/2}}\bigl( \lambda
t\bigr) ^{1-d/2}\int_{0}^{1}ds\ \nonumber\\
&  \times\, \frac{\exp[ -\lambda t\bigl( (1+s)\tau+sq^{2}/(1+s)\bigr) ]
}{(1+s)^{d/2}} \, . \label{Selbst1}%
\end{align}

\begin{figure}[ptb]
\includegraphics[width=8.4cm]{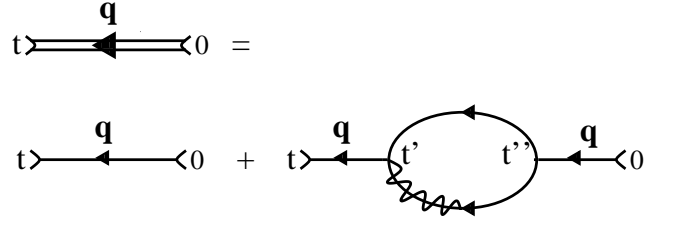}\caption{Dyson equation~(\ref{Dys-Gl})
to 1-loop order.}%
\label{dysonGl}%
\end{figure}

For our purposes we need Green's or connected correlation functions rather
than vertex functions. Hence, we have to consider Feynman diagrams with
external legs attached rather than amputated diagrams. The Green's function
$G_{1,1}$ is determined by the Dyson equation
\begin{align}
G_{1,1}(\mathbf{q},t)  &  =G(\mathbf{q},t)+\int_{0}^{t}dt^{\prime}\int
_{0}^{t^{\prime}}dt^{\prime\prime}\,G(\mathbf{q},t-t^{\prime})\nonumber\\
&  \times\Sigma(\mathbf{q},t^{\prime}-t^{\prime\prime})\,G(\mathbf{q}%
,t^{\prime\prime})+\cdots\,. \label{Dys-Gl}%
\end{align}
A diagrammatic representation of the Dyson equation is given in
Fig.~\ref{dysonGl}.
Upon substituting (\ref{Selbst1}) into (\ref{Dys-Gl}) we obtain after an
integration
\begin{align}
G_{1,1}(\mathbf{q},t)  &  =G(\mathbf{q},t)\biggl[1-\frac{u(\lambda\mu
^{2}t)^{\varepsilon/2}}{\Gamma(1+\varepsilon/2)}\int_{0}^{1}\frac
{ds}{(1+s)^{d/2}}\nonumber\\
&  \times\int_{0}^{1}dx\,(1-x)x^{1-d/2}\exp\bigl(\alpha(s)x\bigr)\biggr]\,.
\label{G1_1}%
\end{align}
Here, we used the shorthand notation
\begin{equation}
\alpha(s)=\Big(\frac{q^{2}}{1+s}-(1+s)\,\tau\Big)\lambda t. \label{alpha}%
\end{equation}
Now, we set $\tau=0$ and expand Eq.~(\ref{G1_1}) to order $q^{2}$. The
integrations are easily performed. After $\varepsilon$-expansion we get%
\begin{align}
G_{1,1}(\mathbf{q},t)  &  =G(\mathbf{q},t)\biggl[1+\frac{u(\lambda\mu
^{2}t)^{\varepsilon/2}}{\Gamma(1+\varepsilon/2)} \bigg( \Big (\frac
{3}{4\varepsilon}+\frac{9}{16}-\frac{\ln2}{8}\Big)\nonumber\\
&  -\Big(\frac{7}{12\varepsilon}-\frac{7}{36}-\frac{\ln2}{24}\Big)\lambda
q^{2}t \bigg) \biggr]\,.
\end{align}
The next step is to remove the $\varepsilon$ poles by employing the
renormalization scheme~(\ref{RenSch}). Letting $G_{1,1}\rightarrow\mathring
{G}_{1,1}$, $\lambda\rightarrow\mathring{\lambda}$, and using
\begin{equation}
G_{1,1}=(\tilde{Z}Z)^{-1/2}\mathring{G}_{1,1}=\Big(1-\frac{3u}{4\varepsilon
}\Big)\mathring{G}_{1,1}%
\end{equation}
as well as
\begin{equation}
\mathring{\lambda}=(\tilde{Z}/Z)^{1/2}\lambda=\Big(1-\frac{7u}{12\varepsilon
}\Big)\lambda\,,
\end{equation}
we observe that the $\varepsilon$ poles are indeed removed. For the
renormalized Green's function we obtain%
\begin{align}
G_{1,1}(\mathbf{q},t)  &  =1+\frac{3u}{8}\Big(\ln(\lambda\mu^{2}t)+C_{E}%
+\frac{3}{2}-\frac{\ln2}{3}\Big)\nonumber\\
&  -\frac{7u}{24}\Big(\ln(\lambda\mu^{2}t)+C_{E}-\frac{2}{3}-\frac{\ln2}%
{7}\Big) \label{G1_1(ende)}%
\end{align}

Two results important for the logarithmic correction can be extracted from
(\ref{G1_1(ende)}). Upon setting $\mathbf{q}=0$ we find
\begin{align}
&  G_{1,1}(\mathbf{q}=0,\lambda\mu^{2}t=X_{0};\tau=0,w;1,1)\nonumber\\
&  =\,1+\frac{3}{8}\Big(\mathcal{Z}+\frac{3}{2}-\frac{\ln2}{3}\Big)w\,,
\label{G1_1(0,0)}%
\end{align}
and hence the amplitude $A_{N}(X_{0})$ as stated in Eq.~(\ref{A_N}). Moreover,
we get
\begin{align}
&  -X_{0}^{-1}\left.  \frac{\partial}{\partial q^{2}}\ln G_{1,1}%
(\mathbf{q},\lambda\mu^{2}t=X_{0};\tau=0,w;1,1)\right\vert _{q^{2}%
=0}\nonumber\\
&  =\,1+\frac{7}{24}\Big(\mathcal{Z}-\frac{2}{3}-\frac{\ln2}{7}\Big)w\,,
\label{G1_1(q,0)}%
\end{align}
which leads to our result for $A_{R}(X_{0})$ given in Eq.~(\ref{A_R(X)}).

\subsection{The Green's function $G_{0,1}$}

\label{app:G01} Now we determine $G_{0,1}$ as required in (\ref{newSurvProb}).
\begin{figure}[ptb]
\includegraphics[width=3.4cm]{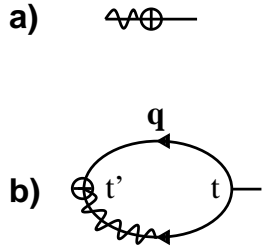}\caption{(a) The new vertex
$-\lambda^{2}g\tilde{M}\,\theta(t-t^{\prime})$ and (b) the 1-loop tadpole
diagram $T(t)$.}%
\label{tadpole}%
\end{figure}
The diagrammatic elements associated with the functional~(\ref{J_k}) comprise
the two vertices encountered in Appendix~\ref{app:G11}. In addition, there is
a third vertex, viz.\ the one depicted in Fig.~\ref{tadpole}a. The Gaussian
propagator for the new functional has to be determined from the differential
equation
\begin{equation}
\Big[\lambda^{-1}\partial_{t}+\tau-g\tilde{M}(t)+q^{2}\Big]\bar{G}%
(\mathbf{q},t,t^{\prime})=\lambda^{-1}\delta(t-t^{\prime})\,. \label{Prop-k}%
\end{equation}
To avoid tadpoles, $\tilde{M}(t)$ has to satisfy the differential equation
\begin{equation}
\overset{.}{\tilde{M}}(t)-\lambda\tau\tilde{M}(t)+\frac{\lambda g}{2}\tilde
{M}(t)^{2}-k(t)+T(t)=0\,. \label{Dgl-shift}%
\end{equation}
At 1-loop order, the tadpole $T(t)$ is given by the diagram shown in
Fig.~\ref{tadpole}b.

The initial and terminal conditions for the fields necessitate the ansatz
$\tilde{M}(t)=-\theta(-t)K(-t)^{-1}$. The type of the source term,
$k(t)=k\delta(t)$ with $k\rightarrow\infty$, demands the initial condition
$K(0)=0$. With this information, the differential equation~(\ref{Dgl-shift})
can be transformed without much effort into the integral equation
\begin{equation}
K(t)+\frac{g}{2\tau}=\mathrm{e}^{\lambda\tau t}\biggl( \int_{0}^{t}
dt^{\prime}\,\mathrm{e}^{-\lambda\tau t^{\prime}}K(t^{\prime})^{2}
T(-t^{\prime})+\frac{g}{2\tau}\biggr) \, . \label{K-Int}%
\end{equation}
At mean field level, the solution to Eq.~(\ref{K-Int}) is given by
\begin{equation}
K_{0}(t)=\frac{g}{2\tau}\Big( \mathrm{e}^{\lambda\tau t}-1\Big) \, .
\label{K_0}%
\end{equation}
Inserting the corresponding $\tilde{M}_{0}(t)=-K_{0}(-t)^{-1}$ into the
differential equation~ (\ref{Prop-k}) we find the modified Gaussian
propagator
\begin{equation}
\bar{G}_{0}(\mathbf{q},t,t^{\prime})=\theta(t-t^{\prime})\biggl( \frac
{K_{0}(-t)}{K_{0}(-t^{\prime})}\biggr) ^{2}\,\exp\Big[ \lambda(\tau
-q^{2})(t-t^{\prime})\Big] \, . \label{Gmod_0}%
\end{equation}

Having the modified Gaussian propagator at our demand, we are now in the
position to calculate the diagram depicted in Fig.~\ref{tadpole}b. Eventually,
we obtain
\begin{align}
&  K(t)^{2}T(-t)=\lambda^{3}g^{2}K_{0}(t)^{-2}\int_{0}^{t}dt^{\prime}%
\int_{t^{\prime}}^{t}dt^{\prime\prime}\nonumber\\
&  \times\,\frac{K_{0}(t^{\prime})K_{0}(t^{\prime\prime})^{2}\exp[\lambda
\tau(2t-t^{\prime}-t^{\prime\prime})]}{[4\pi\lambda(2t-t^{\prime}%
-t^{\prime\prime})]^{d/2}}\,. \label{T_1}%
\end{align}
The further evaluation of Eq.~(\ref{T_1}) is fairly straightforward for
$\tau=0$. Away from criticality, the calculation is more challenging and will be 
addressed
in a future publication~\cite{JaSt_u2}. Here, we find for $\tau=0$ and after
$\varepsilon$ expansion
\begin{equation}
K(t)^{2}T(-t)=-\frac{\lambda g^{3}(\lambda t)^{\varepsilon/2}}{(4\pi)^{d/2}%
}\Big(\frac{5}{8\varepsilon}+\frac{5}{8}-\frac{11\ln2}{16}\Big)\,.
\end{equation}
Insertion of this intermediate result into (\ref{K-Int}) yields
\begin{equation}
K(t)=\frac{g\lambda t}{2}\biggl[1-\frac{u(\lambda\mu^{2}t)^{\varepsilon/2}%
}{\Gamma(1+\varepsilon/2)}\Big(\frac{5}{4\varepsilon}+\frac{5}{8}-\frac
{11\ln2}{8}\Big)\biggr]\,. \label{K-Ergeb}%
\end{equation}
Next, we renormalize. Indicating the consistency of our previous steps, the
appropriate combination of renormalization factors $(Z\tilde{Z}/Z_{u}%
)^{-1/2}=1+5u/(4\varepsilon)+\cdots$ cancels the $\varepsilon$ pole in
(\ref{K-Ergeb}). The renormalized $K(t)$ reads
\begin{equation}
K(t)=\frac{g\lambda t}{2}\biggl[1-\frac{5u}{8}\Big(\ln(\lambda\mu^{2}%
t)+C_{E}+1-\frac{11\ln2}{5}\Big)\biggr]\,.
\end{equation}
Exploiting $G_{0,1}(-t)=K(t)^{-1}$ and $\lambda\mu^{2}t=X_{0}$ as well as
recalling the definition of $\mathcal{Z}$ we finally obtain
\begin{align}
&  \frac{(4\pi)^{3/2}X_{0}}{2}G_{0,1}(-\lambda\mu^{2}t=X_{0};\tau
=0,w;1,1)\nonumber\\
&  =\,w^{-1/2}\biggl[1+\frac{5}{8}\Big(\mathcal{Z}+1-\frac{11\ln2}%
{5}\Big)w\biggr]\,. \label{G0_1(fin)}%
\end{align}



\begin{thebibliography}{99}                                                      
                                         %


\bibitem {BrHa57}S.R.\ Broadbent and J.M.\ Hammersley, Proc.\ Cambridge
Philos. Soc.\ \textbf{53}, 629 (1957).

\bibitem {Mo77}D.\ Mollison, J.\ R.\ Stat.\ Soc.\ B \textbf{39}, 283 (1977).

\bibitem {Li85}T.M. Liggett, \emph{ Interacting Particle Systems} (Springer,
New York, 1985).

\bibitem {Du88}R. Durrett, \emph{Lectures on Particle Systems and Percolation}
(Wadsworth, Pacific Grove, CA, 1988).

\bibitem {Ba85}N.T.J.\ Bailey, \emph{The Mathematical Theory of Infectious
Diseases} (Griffin, London, 1985).

\bibitem {Mu89}J.D.\ Murray, \emph{Mathematical Biology} (Springer, Berlin, 
1989).

\bibitem {ZiGuBa86}R.M.\ Ziff, E.\ Gulari, and Y.\ Barshad,
Phys.\ Rev.\ Lett.\ \textbf{56}, 2553 (1986).

\bibitem {JeFoDi90}I.\ Jensen, H.C.\ Fogedby, and R.\ Dickman, Phys.\ Rev.\ A
\textbf{41}, 3411 (1990).

\bibitem {SchuSe78}L.S.\ Schulman and P.E.\ Seiden,
J.\ Stat.\ Phys.\ \textbf{19}, 293 (1978).

\bibitem {Ja81}H.K.\ Janssen, Z.\ Phys.\ B:\ Cond.\ Mat.\ \textbf{42}, 151 
(1981).

\bibitem {Hi01}H.\ Hinrichsen, Adv.\ Phys.\ \textbf{49}, 815 (2001)

\bibitem {StAh94}D.\ Stauffer and A.\ Aharony, \emph{Introduction to
Percolation Theory} (Taylor \& Francis, London, 1994).

\bibitem {BuHa96}A.\ Bunde and S.\ Havlin (eds.), \emph{Fractals and
Disordered Systems}, (Springer, Berlin, 1995/96).

\bibitem {PaZiSt01}G.\ Paul, R.M.\ Ziff, and H.E.\ Stanley,
Phys.\ Rev.\ E \textbf{64}, 026115 (2001).

\bibitem {Gra02}P.\ Grassberger, Phys.\ Rev.\ E \textbf{67}, 036101 (2003).

\bibitem{osterkamp&Co_2003}
D. Osterkamp, D. Stauffer, and A. Aharony, eprint: cond-mat/0302051.

\bibitem {Gra_pr}P. Grassberger, private communication.

\bibitem {Lue03}S.\ L\"{u}beck, private communication.

\bibitem{essam&Co_78}
J. W. Essam, D. S. Gaunt, and A. J. Guttmann: J. Phys A {\bf 11}, 1983 (1978).

\bibitem{aharony_80}
A. Aharony, Phys. Rev. B, {\bf 22}, 400 (1980).

\bibitem{ruiz-lorenzo_98}
J. J. Ruiz-Lorenzo, J. Phys. A {\bf 31}, 8773 (1998).

\bibitem {GHS94}P.\ Grassberger, R.\ Hegger, and L.\ Sch\"afer, J.\ Phys.\ A
\textbf{27}, 7265 (1994).

\bibitem {GHS99} J.\ Hager, and L.\ Sch\"afer, Phys.\ Rev.\ E
\textbf{60}, 2071 (1999).

\bibitem {JaSt_u1}H.K.\ Janssen and O.\ Stenull, unpublished.

\bibitem {StJa03}O.\ Stenull and H.K.\ Janssen, unpublished.

\bibitem {Gra83}P.\ Grassberger, Math.\ Biosci.\ \textbf{62}, 157 (1983).

\bibitem {Ja85}H.K.\ Janssen, Z.\ Phys.\ B: Cond.\ Mat.\ \textbf{58}, 311 
(1985).

\bibitem {CaGra85}J.L.\ Cardy and P.\ Grassberger, J.\ Phys.\ A:
Math.\ Gen.\ \textbf{18}, L267 (1985).

\bibitem {Ja76}H.K.\ Janssen, Z.\ Phys.\ B: Cond.\ Mat.\ \textbf{23}, 377 
(1976);
R.\ Bausch, H.K.\ Janssen, and H.\ Wagner, \emph{ibid.} \textbf{24}, 113 (1976).

\bibitem {DeDo76}C.\ DeDominicis, J.\ Physique C \textbf{37}, 247 (1976);
C.\ DeDominicis and L.\ Peliti, Phys.\ Rev.\ B \textbf{18}, 353 (1978).

\bibitem {Ja92}H.K.\ Janssen, in: \emph{Dynamical Critical Phenomena and
Related Topics}, Lecture Notes in Physics, Vol.\ 104, ed.\ C.P.\ Enz
(Springer, Heidelberg, 1979); H.K.\ Janssen, in: \emph{From Phase Transition
to Chaos}, ed.\ G.\ Gy\"{o}rgyi, I.\ Kondor, L.\ Sasv\'{a}ri, T.\ T\'{e}l
(World Scientific, Singapore, 1992).

\bibitem {Am84}D.J.\ Amit, \emph{Field Theory, the Renormalization Group, and
Critical Phenomena} (World Scientific, Singapore, 1984).

\bibitem {ZJ96}J.\ Zinn-Justin, \emph{Quantum Field Theory and Critical
Phenomena} (Clarendon, Oxford, 1996).

\bibitem {AKM81}O.F.\ de Alcantara Bonfim, J.E.\ Kirkham and A.J.\ McKane,
J.\ Phys.\ A \textbf{13}, L247 (1980); \emph{ibid.} \textbf{14}, 2391 (1981).

\bibitem{footnote} In Rev.~\cite{Gra02} the number of active sites is denoted by $M$.

\bibitem{footnote2} The field theory of static percolation is founded on the Potts model. The role of the order parameter field is played by a  Potts-spin-like field, say $\phi$. With the notation we use here, $\phi$ is renormalized by the renormalization factor $\tilde{Z}$, viz.\ $\phi \rightarrow
\mathring{\phi}=\tilde{Z}^{1/2}\phi$.

\bibitem {Ja02}H.K.\ Janssen, to be published.

\bibitem {JaSt_u2}H.K.\ Janssen and O.\ Stenull, unpublished.
\end{thebibliography}
\end{document}